\documentclass[apj,numberedappendix]{emulateapj}
\usepackage[latin1]{inputenc}
\usepackage{amssymb}
\usepackage{amsmath}
\usepackage{graphicx}
\usepackage{xspace}
\usepackage{natbib}
\usepackage{url}
\usepackage{paralist}
\usepackage{aasmacros}

\slugcomment{ }

\shorttitle{Fast UV Variability in Fairall~9}
\shortauthors{Lohfink et al.}

\begin{document}

\title{The fast UV variability of the active galactic nucleus in Fairall~9}


\author{Anne M. Lohfink\altaffilmark{1}, Christopher S. Reynolds\altaffilmark{1}, Ranjan Vasudevan\altaffilmark{1}, Richard F. Mushotzky\altaffilmark{1}, Neal A. Miller\altaffilmark{2}}
\email{alohfink@astro.umd.edu}
\altaffiltext{1}{Department of Astronomy, University of Maryland, College Park, MD 20742-2421, USA}
\altaffiltext{2}{Stevenson University, 1525 Greenspring Valley Rd, Stevenson, MD 21153, USA}


\begin{abstract}
We present results from a new optical/UV/X-ray monitoring campaign of the luminous Seyfert galaxy Fairall~9 using the {\it Swift} satellite.  Using the {\it UV-Optical Telescope} (UVOT) on {\it Swift}, we find correlated optical/UV variability on all time scales ranging from the sampling time (4-days) to the length of the campaign (2.5 months).  In one noteworthy event, the UW2-band flux dips by 20\% in 4-days, and then recovers equally quickly; this event is not seen in either the optical or the X-ray bands.  We argue that this event provides further evidence that a significant fraction of the UV-emission must be driven by irradiation/reprocessing of emission from the central disk.  We also use an archival {\it XMM-Newton} observation to examine shorter time scale UV/X-ray variability. We find very rapid ($<\,10$\,ks) UV flares of small amplitude. We show that, unless this emission is non-thermal, we must be seeing the Wien tail from a compact ($<\,3$ light hours), hot ($T>8\times 10^4$\,K) region. The possible association with X- ray microflares suggests that we may be seeing the UV signatures of direct X-ray flare heating of the innermost disk.

\end{abstract}



\section{Introduction}\label{intro}

It is widely accepted that the optical/UV continuum from radio-quiet active galactic nuclei (AGN) is thermal (quasi-blackbody) emission from the radiatively-efficient, optically-thick supermassive black hole accretion disk.  Hence, the spectrum and variability of the optical/UV emission gives us a direct window on the accretion process.  Prior to the launch of {\it Swift} and {\it XMM-Newton}, most studies of the UV variability were performed as part of reverberation mapping campaigns.  For example, in the 1990s, the International AGN Watch Consortium (IAWC) used ground-based optical spectroscopy, as well as UV, EUV and X-ray data from a number of space-based observatories to monitor several Seyfert galaxies including the subject of this paper, Fairall~9 ($z=0.047$).  In a series of works \citep[e.g.][and references therein]{rodriguez:97a,marshall:97a,obrien:98a} the IAWC showed that the optical/UV emissions in these sources were often variable down to the sampling time scale.  Furthermore, they showed that the optical and UV (and sometimes but not always the X-ray) were highly correlated with no detectable time-lag.  Since this implies that emission across an extended region of the disk must be correlated on essentially the light-crossing time, this drove one to a picture whereby the optical/UV emission is reprocessed energy originating from the central-most parts of the accretion disk.  

These results were confirmed by later monitoring campaigns based on X-ray monitoring by the \textit{RXTE} satellite and a ground based optical monitoring. For example, \citet{breedt:09a} studied a five year monitoring of Mrk\,79 determining that the optical and X-ray are correlated without any lag. In NGC\,3516 on the other hand, the results were less conclusive, as absorption does not give us direct view of the central engine during most of the time \citep{maoz:02a}. A two year monitoring of NGC\,3783 by \citet{arevalo:09a} revealed no or a very small lag between the B and V bands, confirming the idea that the UV and optical are originating from adjacent parts of the accretion disk. Between the X-ray band and the UV/optical a 3-9 day lag consistent with the UV being reprocessed emission was detected.

The launches of {\it Swift} and {\it XMM}, with their optical/UV telescopes mounted co-axially to their pointed X-ray instruments have also provided powerful new opportunities to study the optical/UV/X-ray variability of AGN.  Both UV telescopes (the {\it Swift} UV-Optical Telescope [UVOT] and the {\it XMM} Optical Monitor [OM]) have significantly better spatial resolution than IUE, and both observatories  automatically provide simultaneous X-ray observations. These instruments provide advantages over ground-based studies even for optical photometry as they are space-based and hence complications such as airmass can be avoided.  While \textit{XMM} provides the better X-ray spectrum due to its large effective area in the X-ray band, \textit{Swift} is much more suited for monitoring campaigns because its shorter slew time and autonomous operation.   To a large extent, these studies have confirmed the results of IAWC.  For example, \citet{cameron:12a} have presented {\it Swift} monitoring of the low-mass AGN NGC~4395 and find correlated variability from the optical through into the X-ray.   No optical-UV lag is detected, and a very small (400\,s) lag between the X-ray and UV is reported.  

To date, most of the objects studied with this new generation of observatories are blazars (which are completely jet dominated and hence not useful for accretion disk studies), or highly variable objects such as narrow-line Seyfert-1 galaxies \citep{alston:13a, gliozzi:13a, grupe:13a}. Narrow-line Seyfert-1 galaxies are peculiar objects known to possess large soft excesses and mostly accrete at high Eddington rates.  Some results have also been reported for ``normal'' Seyfert galaxies, using these type of instruments.  

The goal of this paper is to shed more light on the UV/optical variability and its relation to the X-ray variability in one of those average Seyfert-like AGN -- Fairall~9. The data used in this analysis are a 2.5-month \textit{Swift} monitoring and an archival \textit{XMM} dataset. Given that this AGN hosts a rather massive black hole, $M=(2.55\pm0.56)\times 10^{8}M_\odot$ \citep{peterson:04a}, we are able  to probe time scales down to almost the light crossing time of the black hole. It is not only its high black hole mass that makes Fairall~9 a suitable study target, but it is also unobstructed by any line-of-sight absorption such as warm absorbers \citep[e.g.][]{emmanoulopoulos:11a, lohfink:12a}. This lack of line-of-sight absorption offers the possibility to study the continuum and its variability in great detail and excludes absorption as the driver of the variability in this object as it has been previously suggested in some cases \citep[e.g.,][]{grupe:13a}.   

We discover UV variability on 4\,day time scales,  the light crossing time of the UV emitting region; we argue that this is strong confirmation that a significant component of the UV is reprocessing of (largely unseen) EUV emission from the central parts of the accretion disk. The \textit{XMM} dataset allows us to study the UV and X-ray variability on the time scales of a few hours. We calculate the Edelson \& Krolik discrete correlation function and find two marginally significant correlations. One is a correlation were the UV lags the X-rays by about 1-2\,hours.  The second is an anti-correlation where again the UV lags the X-rays, this time by about 5\,hrs. The 1-2\,hrs lag can be explained by reprocessing of the X-ray emission of small, short duration X-ray flares in the accretion disk. The anti-correlation however is hard to explain and will require further investigation.    

The outline of this paper is as follows. First, we describe the datasets used in this work and briefly discuss data reduction techniques (\S \ref{data}). After an investigation of the optical/UV variability (\S \ref{uv_variability}), we analyze the UV--X-ray variability (\S \ref{uvxray_variability}).   We end with a  discussion of the implications of these results (\S \ref{discussion}).

\section{Data Reduction}\label{data}
\subsection{Basic \textit{Swift} data reduction}
The {\it Swift} data presented here is from a 2.5 month campaign from 16-April-2013 to 1-July-2013 designed to search for and further study the X-ray dips first reported in \citet{lohfink:12a}.   The XRT data were taken in windowed timing mode and uniformly reduced and analyzed as described in \citet{lohfink:13a}.

UVOT was operated in imaging mode and, for most of the pointings, cycled through all six filters;  V, B, U, UW1, UM2, and UW2. The UVOT analysis begins with summing the individual exposures per observation for each individual UVOT filter using the tool \texttt{uvotimsum}. The \texttt{uvotsource} tool was then used to extract fluxes from the images using aperture photometry, it also corrects for coincidence losses. The source region was selected to be circular with a 5.0\,arcsec radius around the coordinates obtained from the NASA Extragalactic Database (NED) ---RA 01h23m45.8s, DEC $-$58d48m20.8s (J2000). For the background extraction, a source-free circular region with 32 arcsec radius close to Fairall~9 was selected. 

\subsection{Basic \textit{XMM} data reduction}

The archival \textit{XMM-Newton} dataset considered here was taken 10-December-2009 (ObsID:0605800401). The X-ray data were reduced as described in \citet{lohfink:12b}. We only consider EPIC-pn data for the analysis performed in this paper, as MOS does not contribute much additional information for this work.  The total good on-source exposure is 91\,ks.   From this, we extract a lightcurve binned to a time resolution of 1\,ks. The \textit{XMM}-OM was operated with a single UV-filter (UVW1), and the data were reduced using the \texttt{omipipeline}.  The resulting count rates where then converted into fluxes as outlined in the XMMSAS User's Guide \footnote{http://xmm.esac.esa.int/external/xmm\_user\_support/documentation/sas\_usg/USG/ommag.html}.

\subsection{Differential Photometry}

\begin{figure}
\includegraphics[width=0.96\columnwidth]{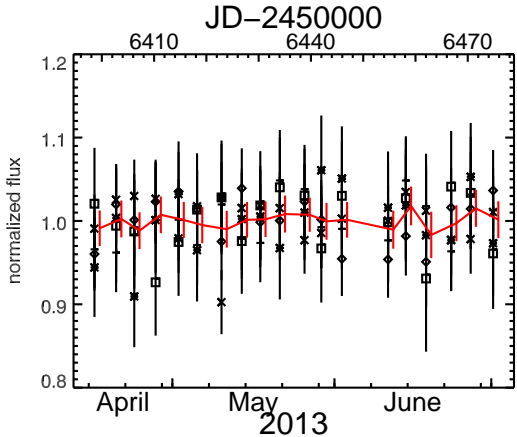}
\caption{Normalized fluxes of reference stars, where each plot symbol corresponds to a different star, and their average flux in U-Band shown in red color. The average, normalized flux is offset in time in the plot for clarity.}\label{guide}
\end{figure}

To enhance the accuracy of the UV/optical photometry and verify the photometric stability during the 2.5 months of monitoring, we perform differential photometry with respect to reference stars in the vicinity of Fairall~9. We consider a total of 6 stars, which were selected from the USNO-A2 guide star catalog. A list of those stars is given in Table~\ref{stars}. These stars are detectable in every single frame image in all the filters for the UVOT instrument.  We then proceed to obtain the fluxes in all filters of all these stars, in case of UVOT, and in case of OM of stars \# 1, 2 and 4. For UVOT this is done by again using a 5\,arcsec source radius and making use of the tool \texttt{uvotsource}.  For OM the photometry results from \texttt{omdetect} can be used directly. The stars' average observed B-V color is 0.01, which is somewhat bluer than that of Fairall~9 (0.13) and most likely arises from the requirement of a detection also in the bluer optical bands. Assuming the stars do not vary on the relevant time scales, the attained individual flux values are re-normalized and averaged for each observation. An example of the resulting reference star light curves and its average can be seen in Figure~\ref{guide} for the UVOT U-Band. These re-normalized, averaged values now represent the variability due to the detector or other observation related differences. To correct our Fairall 9 flux values by this variability and get the true variability, we divide by the variability seen in the reference stars. 

\begin{table}
\caption{Reference stars selected to monitor the detector variability of \textit{Swift}-UVOT}\label{stars}
\begin{tabular}{c|c|c|c|c}
Source \# &USNO A02 identifier & Ra & Dec & Distance \\ 
 & & & & [arcmin]\\
 \hline \hline
1 & 0300-00420304 &	20.944 & -58.795 & 0.62 \\
2 & 0300-00421798 & 21.007 & -58.829 & 2.51 \\
3 & 0300-00422565 & 21.038 & -58.730 & 5.46 \\
4 & 0300-00419203 & 20.894 & -58.889 & 5.20 \\
5 & 0300-00423563 &	21.081 & -58.732 & 6.23 \\
6 & 0300-00419618 &	20.913 & -58.910 & 6.33 \\\hline
\end{tabular}
\end{table}

\subsection{Host Galaxy Subtraction}
The detector-variability corrected UV fluxes are corrected for reddening from our Galaxy using a reddening law by \citet{cardelli:89a} with updates in the near-UV by \citet{odonnell:94a}, assuming E(B-V)=0.03 \citep{schlegel:98a}. However, the measured flux is not all just nuclear/AGN emission --- to obtain the optical/UV lightcurve of the AGN, this lightcurve needs to be corrected for contamination by the host galaxy.   In particular, the (time-invariant) host galaxy component can offset the spectral slope and mimic a slope change which is not real. Therefore, a proper host galaxy correction is necessary if we are to study color changes and variability amplitudes in the AGN with any degree of robustness.

\begin{figure*}[htb]
\begin{center}
\includegraphics[width=0.7\textwidth]{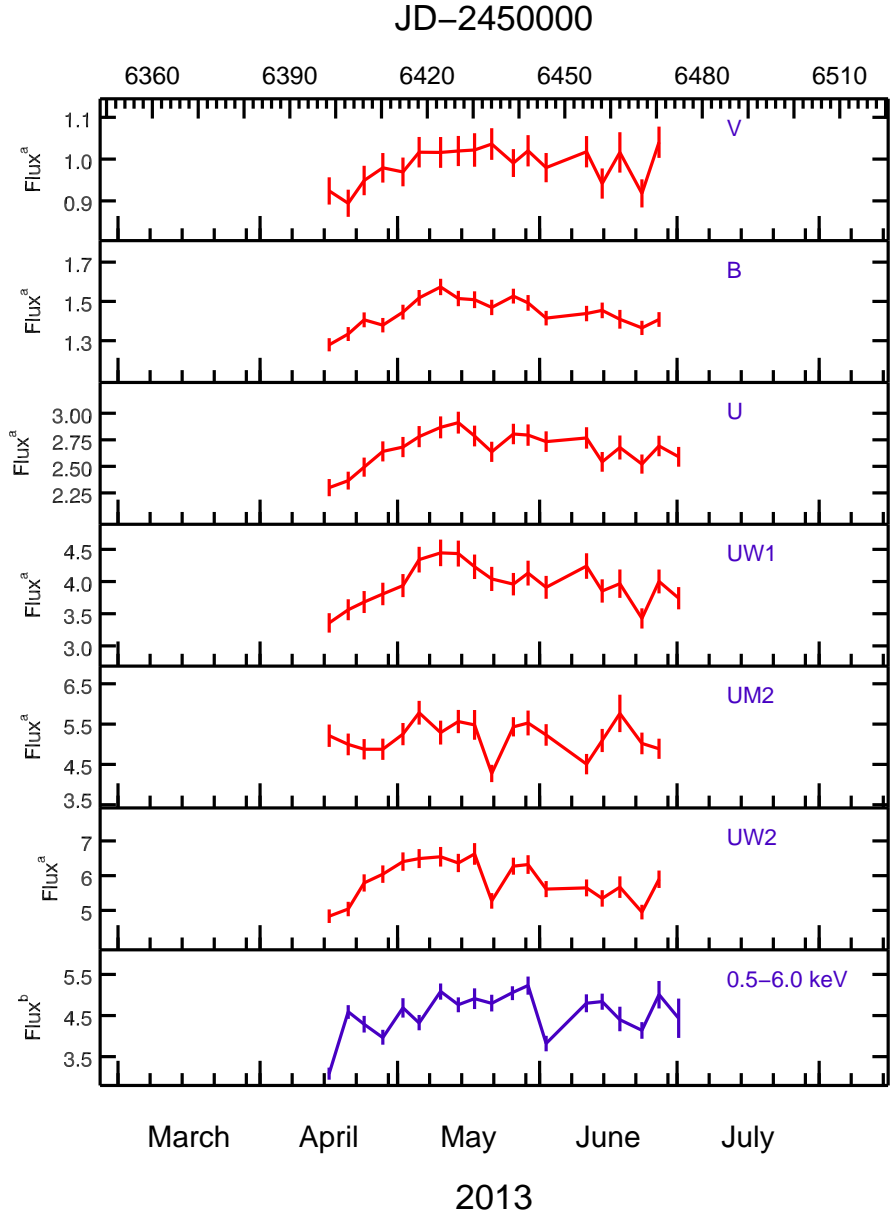}
\caption{Fairall~9 \textit{Swift} lightcurve, with an average 4-day sampling, at selected wavelengths corrected for guide star variability [blue], as well as reddening and host galaxy contribution [red]. $^a$ Flux in $10^{-14}\,\text{erg}\,\text{s}^{-1}\,\text{cm}^{-2}\,\text{\AA}^{-1}$. $^b$ Flux in $10^{-11}\,\text{erg}\,\text{s}^{-1}\,\text{cm}^{-2}\,\text{keV}^{-1}$}\label{f9_uv}
\end{center}
\end{figure*}

\begin{figure}[h]
\includegraphics[width=\columnwidth]{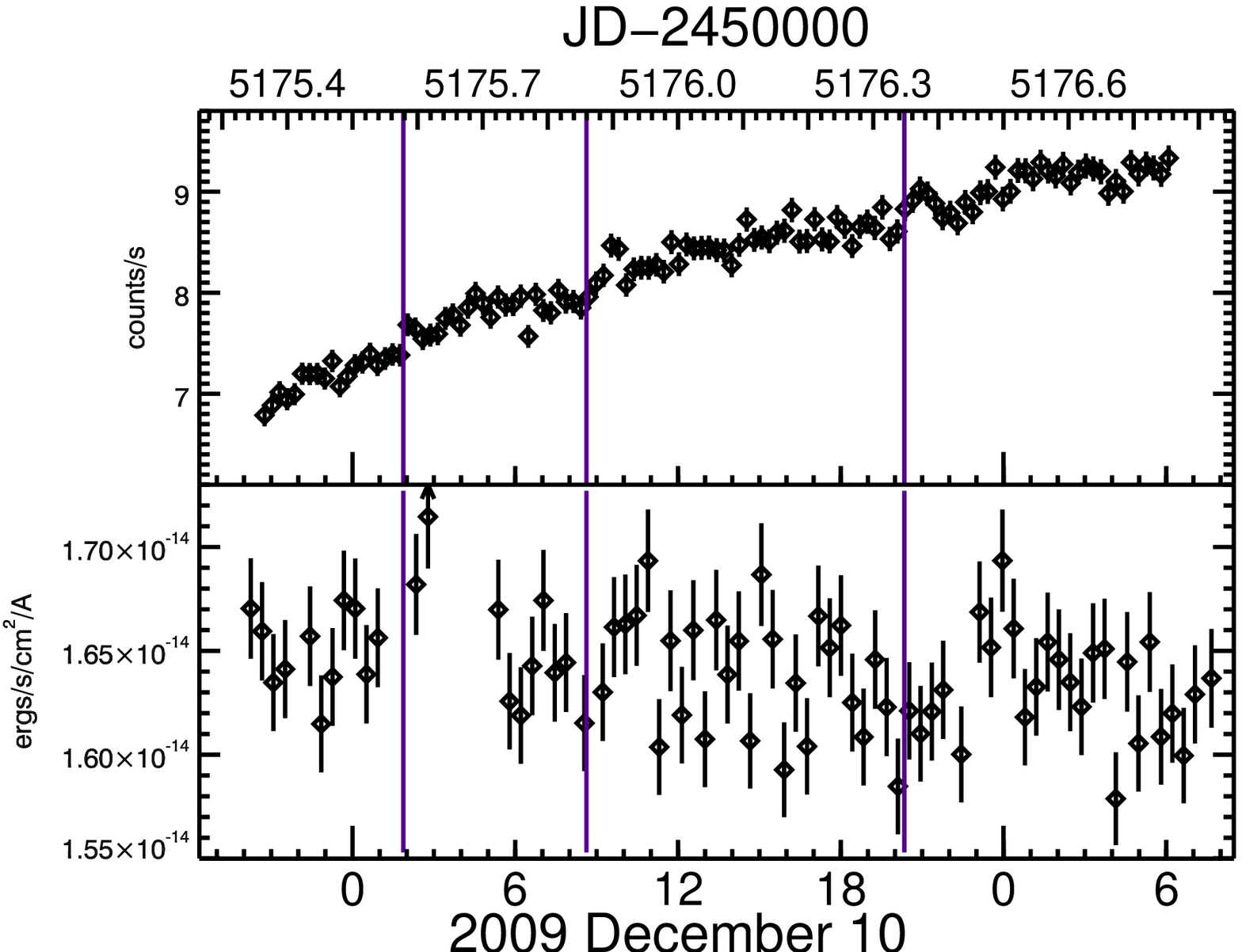}
\caption{Fairall~9 \textit{XMM} X-ray lightcurve (top panel) [0.3-2.0 keV, 1\,ks resolution] and UV fluxes (bottom panel) corrected for guide star variability, Galactic reddening and the host galaxy [2910\,\AA, $\sim$\,1.8\,ks resolution].}\label{uv_xmm}
\end{figure}

The host galaxy contribution to the flux at optical wavelengths was studied with HST by \citet{bentz:09a} and the galaxy classified as a SBa galaxy. By performing a two-dimensional galaxy decompositions they determine the host galaxy flux of Fairall~9 at 5100\,\AA\, to be $3.47\times10^{-15}\,\text{erg}\,\text{s}^{-1}\,\text{cm}^{-2}\,\text{\AA}^{-1}$.  52\,\% of this galaxy flux originated from the bulge.  As the \textit{Swift}-UVOT and \textit{XMM}-OM have a very limited spatial resolution compared to HST, we cannot perform a detailed study of the host galaxy ourselves, but instead use the HST value to anchor SED templates which can then be used to determine the galaxy contribution to our UVOT/OM light-curves. Galaxy SED templates are preferable as they enable the host galaxy estimation at various wavelengths, while galaxy decompositions are only meaningful at the wavelength performed. The region used for the UVOT photometry is small so we assume that all of the galaxy flux in this central region comes from the bulge. We also verify that this decision, which underestimates the galaxy contribution slightly, yields qualitatively the same results for variability and color changes than overestimating galaxy contribution by subtracting off the entire galaxy (bulge+disk). We start the galaxy subtraction by renormalizing the bulge template from \citet{kinney:96a} to the bulge flux measured with HST. The host galaxy flux contribution in a given UVOT/OM filter band $F_\text{filter}$ can then be obtained by convolving the normalized filter function $S_\text{norm}(\lambda)$ with the re-normalized bulge template $T_\text{norm}(\lambda)$
\begin{equation}  
F_\text{filter}=\int T_\text{norm}(\lambda)\,S_\text{norm}(\lambda) d\lambda 
\end{equation}
The so-calculated galaxy fluxes are subtracted from the measured fluxes and we are able to obtain the final UV/optical light curves (Fig.\,\ref{f9_uv} and Fig.~\ref{uv_xmm} [bottom panel]).  In all bands, the reddening correction dominates (and grows in importance into the UV).  The host galaxy correction is modest in the B-band ($\sim 6$\%), and decreases in importance at shorter wavelengths.  

\begin{table*}
\caption{Mean AGN fluxes, Galaxy fluxes and normalized excess variances per filter band/center filter wavelength.}\label{fractions}
\begin{center}
\begin{tabular}{c|c|c|c|c}
Band & $\lambda_\mathrm{c}$ & $\bar{F}_\mathrm{AGN}$ & $F_\mathrm{Gal}$ & $\sigma^2_\mathrm{rms}$ \\ 
\hline
 & \AA & $10^{-14}$\,$\text{erg}\,\text{s}^{-1}\,\text{cm}^{-2}\,\text{\AA}^{-1}$ & $10^{-16}$\,$\text{erg}\,\text{s}^{-1}\,\text{cm}^{-2}\,\text{\AA}^{-1}$ & $10^{-3}$ \\
 \hline \hline
V & 5468 & 0.959 & 15.9 & $3.6\pm1.1$ \\
B & 4392 & 1.39 & 8.93 & $3.9\pm0.86$ \\
U & 3465 & 2.56 & 3.10 & $5.3\pm1.3$ \\
UW1 & 2600 & 3.76 &	0.853 & $8.9\pm2.2$ \\
UM2 & 2246 & 4.82 & 0.460 & $10\pm3.1$ \\
UW2 & 1928 & 5.49 & 0.601 & $12\pm2.6$ \\\hline
\end{tabular}
\end{center}
\end{table*}

\section{Results}
\subsection{Optical/UV Variability}\label{uv_variability}

Figure~\ref{f9_uv} shows the optical/UV/X-ray light curves from our {\it Swift} monitoring and indicates variability in all bands. This is verified when performing a $\chi^2$-test, variability is detected in each band with at least 95\,\% confidence. The variability increases in amplitude towards the UV bands (Table~\ref{fractions}) and is clearly correlated between the different bands. For example, we find a correlation coefficient of 0.64 (p-value: 0.004) between the V- and the UW2-band, neglecting the small delay between the observations originating from the filter rotation.  

It is interesting that we detect UV variability down to the shortest time scales probed by the \textit{Swift} campaign (4\,days). Particularly noteworthy is the UV dip on 22-May-2013.  During this event the UW2 (1928\AA) flux drops by 20\% between two pointings separated by 4-days.  This dip is also seen in the UM2 filter (which possesses a band-pass that overlaps significantly with the UW2), but is essentially absent in the optical bands.  

\begin{figure}[h]
\includegraphics[width=0.96\columnwidth]{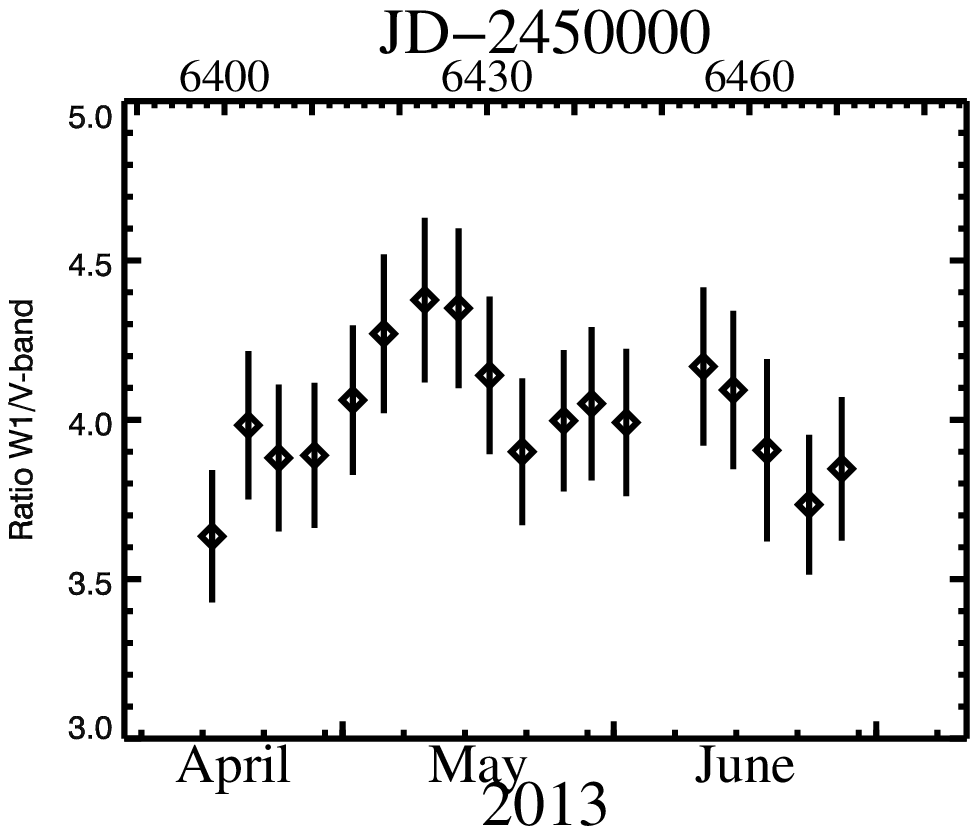}
\caption{UV hardness evolution between the UW1 and V band during the \textit{Swift} monitoring.}\label{uv_hard}
\end{figure}

The next question is whether this change in optical/UV flux is accompanied by a change of spectral slope in the optical/UV. Previous analyses of the optical variability have found Fairall~9 to show peculiar spectral behavior in the UV/optical, not hardening when brightening \citep{recondo:97a,santos:97a}, as it is observed in most sources \citep[e.g.,][]{VandenBerk:04a}. We therefore investigate whether our pure AGN fluxes show any signs for spectral variability between the optical and UV bands. 
Figure~\ref{uv_hard} shows the flux ratio between the UW1-band flux and the V-band flux. Changes in color are clearly detected, with the UV/optical becoming harder when brighter, opposite to the trend observed in the X-ray band \citep{emmanoulopoulos:11a}.  The average, measured power law slope during the monitoring was 1.33$_{-0.02}^{+0.02}$. The steepest slope measured was 1.48$_{-0.06}^{+0.07}$ and the flattest 1.25$_{-0.06}^{+0.06}$. The average slope is substantially steeper than the average quasar slope determined by \citet{davis:07a} from the SDSS. While this discrepancy could generally be explained by intrinsic reddening, this is unlikely in case of Fairall~9 as \citet{crenshaw:01a} determined from HST observations it does not possess intrinsic reddening. Our detection of color variation is, on the face of it, at odds with previous studies of this source \citep{recondo:97a,santos:97a}. The primary culprit appears to be the host galaxy subtraction. \citet{santos:97a} essentially assumed that the AGN displayed achromatic variability and used this fact to define a galaxy subtraction. Our analysis appears to suggest that a more realistic galaxy subtraction does indeed result in AGN color changes.       

The {\it XMM-Newton} data considered here give us a view of the UV variability of Fairall~9 on time scales of hours.   As can be see in Fig.~\ref{uv_xmm} (bottom panel), the general trend seems to be a flux decline over the length of the observation ($\sim$\,1\,day).  However, there are a few short and small amplitude flares observed, the beginning of each is marked on Fig.~\ref{uv_xmm}  by a vertical line.  These flares last approximately 10\,ks ($\sim 3$\,hours) with an amplitude approximately 2--4\% above the quiescent level, i.e., monochromatic flare luminosities of $\Delta L(2920\AA)=1-2\times 10^{39}\,{\rm erg}\,{\rm s}^{-1}\,\AA^{-1}$ (although the first flare occurred just as the OM entered a short period of inoperation and may have been brighter).

\subsection{X-ray variability and the X-ray/UV connection}\label{uvxray_variability}

After investigating the UV variability we now turn to the X-ray variability of the source as well as the relationship between the X-ray and UV variability.

\begin{figure}
\includegraphics[width=0.96\columnwidth]{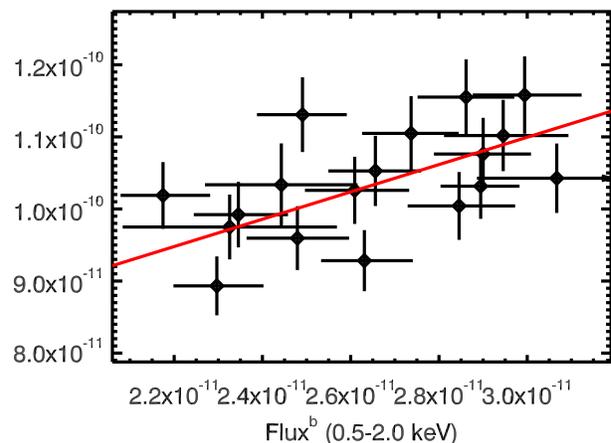}
\caption{The soft, absorbed X-ray (0.5-2.0\,keV) flux of Fairall~9 versus its de-reddened UV (2600\,\AA) flux, a positive correlation is apparent (Pearson coefficient: 0.66 , p-value:  0.002). The red solid line represents the best linear fit. $^b$Flux in $\text{erg}\,\text{s}^{-1}\,\text{cm}^{-2}$.}\label{uvxray_swift}
\end{figure}

On the longer time scales probed by our {\it Swift} monitoring campaign, high amplitude X-ray variability (40\%) is clearly seen in the 0.5--6\,keV band (Figure~\ref{f9_uv}, bottom panel).  The general trend is a weak positive correlation between the UV and X-ray bands with a Pearson coefficient of 0.66 and a p-value of 0.002 (Fig.~\ref{uvxray_swift}). The value for the linear slope is $1.90\pm 0.37$. Given the lack of a detailed correspondence between any given optical/UV lightcurve and the X-ray lightcurve, as well as the fact that this monitoring campaign only {\it just} probes down to the break in the power density spectrum ($f_b\approx 4\times 10^{-7}\approx 1\,{\rm month}^{-1}$; \citet{markowitz:03a}) it is not possible to say from these data to what degree the 0.5--6\,keV X-ray flux is truly related with the optical/UV band on these time scales.  

As we discuss in more detail in Section~\ref{discussion}, it is noteworthy that the strong UV-dip on 22-May-2013 is not reflected in the gross X-ray lightcurve.  On the other hand, two weeks later (3-June-2013), there is a significant X-ray dip that has no correspondence in any of the optical or UV bands (unfortunately, the observing campaign was briefly interrupted in early June-2013 due to a higher-priority {\it Swift} target of opportunity, and hence we did not witness egress from this X-ray dip).

\begin{figure}[h]
\includegraphics[width=0.96\columnwidth]{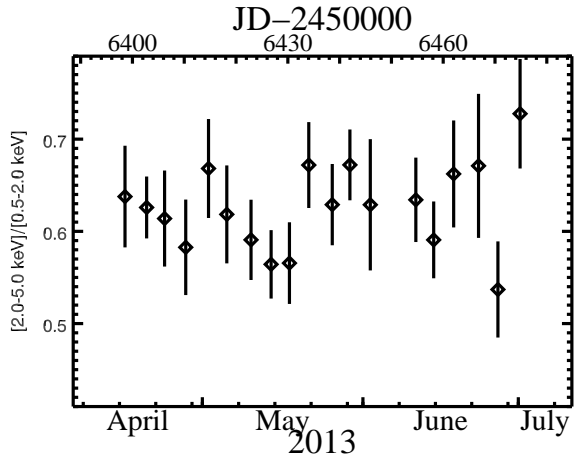}
\caption{X-ray hardness evolution between the 2-5\,keV flux and the 0.5-2\,keV flux during the \textit{Swift} monitoring.  }\label{hardness_xray}
\end{figure}

However, the UV-dip may be correspond to an X-ray spectral change.  Each individual XRT pointing has insufficient counts for detailed spectroscopy, but a crude search for X-ray spectral variability can be conducted using hardness ratios.  We choose to examine the ratio of the 0.5--2\,keV and 2--5\,keV fluxes (derived from the corresponding count spectra with a spectral model consisting of a power-law continuum modified by Galactic absorption, $N_H=3.2\times 10^{20}\,{\rm cm}^{-2}$).   As shown in Fig.~\ref{hardness_xray}, the X-ray hardness ratio is approximately constant but there is marginal evidence for a rapid hardening at exactly the same time as the UV emission dips.

\begin{figure}[h]
\includegraphics[width=0.96\columnwidth]{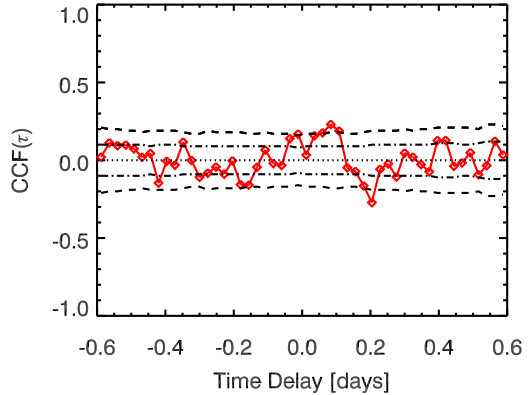}
\caption{DCF for \textit{XMM} X-ray and UV lightcurve (red line). The dot-dashed and dashed lines represent the $1\,\sigma$ and $2\,\sigma$ confidence limits, respectively. In the plot a peak at positive lag times indicates that the UV is lagging the X-rays by this time.}\label{ccf_xmm}
\end{figure}

More rapid X-ray/UV connections can be explored using the {\it XMM} data.  The X-ray variability within this pointing was already discussed in \citet{emmanoulopoulos:11a}; they find a gradual softening of the X-ray spectrum, while the flux continuously increases.   While there is generally little short term (hour-time scale) variability, we do see small flares that last approximately 2--4\,ks with approximately 5\% amplitude (Fig.~\ref{uv_xmm}).   Interestingly, from examination of Fig.~\ref{uv_xmm}, we discover hints that the short-term flares are not random but it appears that the X-ray is leading the UV by a 1-3\,hours. The three most apparent flares are marked in the lightcurve in Fig.~\ref{uv_xmm}. To confirm this apparent time lag we calculate the Edelson \& Krolik discrete correlation function (DCF) \citep{edelson:88a} for the \textit{XMM}-pn and \textit{XMM}-OM light curves after pre-whitening them with a second order polynomial \citep{welsh:99a}. The resulting DCF is shown Fig.~\ref{ccf_xmm}, with $1\,\sigma$ and $2\,\sigma$ confidence levels estimated by model-independent Monte-Carlo simulations. With our interest being focused on the existence of correlations, we simulate 5000 DCFs to estimate the robustness of our potential lags, following these steps: 

\begin{itemize}
\item[Step 1)] Make a synthetic UV and X-ray lightcurve, by drawing, for each timestamp in the lightcurve, a flux/count rate value from a Gaussian distribution defined by the datapoint from the pre-whitened lightcurves as its mean and the errorbar as its sigma, 
\item[Step 2)] Randomize the values within each synthetic lightcurve but keep the timestamps fixed, 
\item[Step 3)] Calculate the DCF between the two lightcurves using the same lag bin values as in the actual DCF, 
\item[Step 4)] After repeating Steps 1-3 5000 times, estimate the confidence levels.
\end{itemize} 
These simulations estimate the statistical errors on the DCF, as well as the strength of the correlations found, by comparing them to how likely it is to get correlations of such from randomized light curves. A positive lag in the Figure~\ref{ccf_xmm} means the UV is lagging behind the X-rays, i.e. the X-ray variations leading the UV variations. There is no correlation or time lag detected with great confidence, we note, a peak at a UV lag time of about 1-2\,hours and a potential anti-correlation on a 5\,hour time scale. Both peaks are on the $2\,\sigma$ level.   

\section{Discussion}\label{discussion}

Our most curious finding is the existence and nature of the rapid UV variability.  Here, we discuss the implications of these observations for our understanding of the UV continuum in this AGN.

In the standard picture, the optical/UV continuum is thermal (quasi-blackbody) radiation from the optically-thick accretion disk.  Adopting the standard model for a geometrically-thin, radiatively-efficient, mass-conserving accretion disk by \citet{shakura:73a}, the effective temperature profile is given by
\begin{equation}
T_{\rm eff}=1.08\times 10^6\,{\cal L}^{1/4}M_8^{-1/4}\eta_{0.1}^{-1/4}r^{-3/4}f\,{\rm K},
\end{equation}
where ${\cal L}=L/L_{\rm Edd}$ is the Eddington ratio assuming a radiative efficiency of $\eta=0.1\eta_{0.1}$, the black hole has mass $M=10^8M_8M_\odot$, and $r$ is the radius normalized to gravitational units, $r=R/R_g$, $R_g=GM/c^2$.  We have also included a factor $f=[1-(r_{\rm isco}/r)^{1/2}]^{1/4}$ which encapsulates the affect of the zero-torque boundary condition at the innermost stable circular orbit, $r=r_{\rm isco}$.   Strictly, this form for the temperature profile becomes inaccurate in the innermost disk ($r\lesssim 10$) due to the emergence of relativistic corrections.  For Fairall~9, we take $M_8=2.55$ \citep{peterson:04a} and ${\cal L}=0.15$ \citep{lohfink:12a} resulting in a temperature profile,
\begin{equation}
T_{\rm eff}=5.3\times 10^5\eta_{0.1}^{-1/4}r^{-3/4}f\,{\rm K}.
\end{equation}

The half-light radius of the thermal emission at some particular wavelength is determined by integrating the thermal emission outwards from the innermost radius of the disk (which we shall take to be $r=6$).  Employing the color-correction factors described by equations~(1) and (2) of \cite{done:12a}, we find that the half-light radius for the UW2-band emission (with central wavelength $\lambda=1928$\AA\ and corresponding blackbody temperature $T\approx 1.5\times 10^4$\,K) is $r_{1/2}\approx 52$.  Similarly, the half-light radius of the B-band emission (with central wavelength $\lambda=4392$\AA\ and corresponding blackbody temperature $T\approx 6.6\times 10^3$\,K) is $r_{1/2}\approx 118$.   The corresponding light crossing time of the UV (optical) region is $t_{\rm lc}=2r_{1/2}/c\approx 1.5$\,days (3.5\,days), the dynamical timescale is $t_{\rm dyn}\equiv (R^3/GM)^{1/2}=5.5$\,days (19\,days), the thermal time scale is $t_{\rm th}=t_{\rm dyn}/\alpha=55$\,days (190\,days) assuming a Shakura-Sunyaev viscosity parameter of $\alpha=0.1$, and the viscous time scale is $t_{\rm visc}=t_{\rm th}/(h/r)^2=1.5\times 10^3$\,yr ($5.2\times 10^3$\,yr) assuming a disk aspect ratio of $h/r=0.01$. A summary of the properties of the optical emission region and the just calculated time scales can be found in Table~\ref{timescales}.

\begin{table}
\caption{Key properties of the optical UV/emission regions (for the B and UW2 band) and the corresponding charateristic time scales. A detailed explanation can be found in the text.}\label{timescales}
\begin{tabular}{l|c|c}
 Filter Band & B & UW2 \\
 \hline 
Center wavelength [\AA] & 4392 & 1928\\
Temperature [K] &	$6.6\times 10^3$ & $1.5\times10^4$ \\
Half light radius [$r_\mathrm{g}$] & 118 & 52 \\
 \hline \hline
light crossing time scale [days] & 3.5 & 1.5\\
dynamical time scale [days] & 19 & 5.5\\
thermal time scale [days] & 190 & 55  \\
viscous time scale [years] & $5.2\times10^3$ & $1.5\times10^3$ \\
\end{tabular}
\end{table}

Most of these time scales are considerably longer than the UV variability reported in this paper observed and it therefore cannot simply be attributed to accretion rate changes or thermal instabilities in the UV emitting region of a standard disk. These possibilities are also unlikely considering the fractional variance of only 6\,\% (V-Band) to 11\,\% (W2-Band), implying that only a small part of the total UV/optical emission is actually variable. That a harder when brighter trend is still observed (as it is in our monitoring) even when the accretion rate is not the driver has already been noted by \citet{ruan:14a}. With the nature of the variability being uncertain, the observed UV changes demand further discussion. 

We will begin by addressing the UV variability observed in the \textit{Swift} monitoring. The smooth variations seen in the UV are weakly correlated with the variations seen in the soft X-ray band. This correlation could be explained by two scenarios: either a reprocessing scenario where the X-rays are heating the disk and causing an increase UV/optical emission; or by the upscattering of the UV/optical photons into X-ray band, as expected to produce the observed X-ray power law. Both scenarios would lead to a delay between the two bands, in case of Comptonization the UV would be leading the X-ray variability while reprocessing would imply a lag of the UV behind the X-ray band. The shortness of the monitoring however prevents the measurement or even detection of any such delay. 

The 4-day UV dip seen in our {\it Swift} monitoring requires special attention.  As discussed in Section~\ref{uv_variability}, this dip is seen strongly in the UV filters, but not in the optical bands or the (total) X-ray band.   Given the strong wavelength dependence of dust extinction, an initially appealing hypothesis is an eclipse of the UV emitting region by a wisp of dusty gas in the circumnuclear environment.  However, the short time scales are extremely problematic for this scenario.  Given that the UV emitting region should be at least 3.5 light days in diameter, the wisp would need to be moving at relativistic speeds in order to block, and then unblock, the UV source on the observed time scales.  Any dusty gas will be confined to beyond the sublimation radius, with corresponding velocities that are at least two orders of magnitude smaller.   Thus, we conclude that the UV dip must correspond to a true decrease in the UV emission rather than an absorption event.

If we hold onto the notion that the UV emission has a thermal origin, we see that the observed UV variability occurs on the light crossing time scale.  This immediately leads again to a picture whereby a significant fraction of the UV emission is reprocessed energy from the central most regions of the disk or the corona, and the dip corresponds to an abrupt decrease of incident emission from the central disk or corona.  Since the X-ray emission remains rather steady during the dip, contrary to what is generally observed, there is no evidence that the central accretion disk emission actually shut off or the coronal emission decreased.  Instead, it is possible that a slight change in geometry of the central disk (such as the creation of a transitory equatorial wind) resulted in the shielding of the UV emitting region from the central radiation field and thereby causing this particular type of variability.   

The general, anti-correlated flux trends observed in the X-ray and the UV bands during the \textit{XMM} pointing are consistent with reprocessing being the major driver of variability on the time scale of days to a week, as suggested by the \textit{Swift} monitoring. This is because the pointing length of 91\,ks ($\sim\,1.05$\,days) is shorter than the expected delay between the two bands. Thermal reprocessing of X-ray emission in the disk has also recently been found to drive the short-term UV variability in NGC\,4051 \citep{alston:13a} and PG\,1211+143 \citep{bachev:09a}.

On the other hand the very rapid UV flares seen in the {\it XMM} data are enigmatic.  Their rapid time scales and potential association with X-ray flares strongly argues that these UV flares originate from the centralmost regions of the around the black hole.  Retaining the assumption that this is still thermal/quasi-blackbody emission, we can use the luminosity and duration of the flares to constrain the temperature of the emitting region.  Consider a planar region of area $\pi R^2$ and temperature $T$ observed at an angle $\theta$.  The monochromatic luminosity at wavelength $\lambda$ is
\begin{equation}
L_\lambda=\frac{2\pi R^2\cos\theta hc^2}{\lambda^5(e^{hc/k_BT\lambda}-1)}.
\end{equation}
Thus, if we have a flare of amplitude $\Delta L_\lambda$ and duration $\Delta t$, we can use the fact that $R<c\Delta t/\sin\theta$ to deduce that a causal flare requires
\begin{equation}
\Delta t>\left[\frac{\lambda^5\Delta L_\lambda(e^{hc/k_BT\lambda}-1)\sin^2\theta}{2\pi hc^4\cos\theta}\right]^{1/2}.
\end{equation}
Applying this constraint to the rapid UV flares in Fairall~9, we set $\lambda=2910$\AA, $\Delta t=10$\,ks, $\Delta L_\lambda=2.2\times 10^{39}\,{\rm erg}\,{\rm s}^{-1}\,{\rm \AA}^{-1}$ (corresponding to $\Delta F_{\lambda}=5\times 10^{-16}\,{\rm erg}\,{\rm s}^{-1}\,{\rm cm}^{-2}{\rm \AA}^{-1}$ at a luminosity distance of 193\,Mpc), and $\theta=48^\circ$ \citep{lohfink:13a}.   We conclude that the temperature of the flare region must be $T>8\times 10^4$\,K.  This is comparable to the expected disk temperatures within $r=15$.  

The time scales of these flares (comparable to the light crossing time of $10R_g$) again suggests a reprocessing model.  At first sight, it is tempting to consider that the UV flares originate from heating of the central accretion disk by the observed X-ray micro-flares.  Our tentative detection of a time delay between the X-ray flares and the UV flares could then be attributed to a reverberation delay provided that the X-ray source is only a few gravitational radii from the disk.   However, the energetics of this picture need some consideration.   Extrapolating across all wavelengths, the luminosity of the UV flares will be approximately $2\times 10^{44}\,{\rm erg}\,{\rm s}^{-1}$, peaking in the extreme-UV ($\lambda_{\rm peak}\approx 360$\AA).  On the other hand, the luminosity of the observed X-ray micro-flares  is only $8\times 10^{42}\,{\rm erg}\,{\rm s}^{-1}$ even when extrapolated from 0.02--100\,keV assuming a $\Gamma=2$ powerlaw.  Thus, the X-ray flare heating scenario for the UV flares requires either an additional and very strong soft ($<0.3$\,keV) component to the driving flare, and/or strong beaming of the flare emission towards the disk.  Such beaming could result from relativistic streaming motions of plasma towards the disk in a magnetic flare-loop \citep{field:93a,reynolds:97c} or gravitational light bending in the strong gravity of the black hole \citep{martocchia:96a,reynolds:97b,miniutti:04a}. Alternatively, a non-thermal origin is also a possibility for the observed microvariability.  

The anti-correlation seen in DCF with the X-rays leading the UV by about 5\,hours is very puzzling. The only other reported anti-correlation has been found in NGC\,7469 by \citet{nandra:98a} from a 30\,day RXTE/IUE monitoring. The determined lag time was 4\,days and most likely, if real, this can attributed to complex absorption in the source. With Fairall~9 being unobscured and the lag only 5\, hours, this anti-correlation, if it can be substantiated with future observations, would pose serious questions to our current understanding of the geometry and processes in the vicinity of the black hole.  

\section{Conclusions}

Using data from {\it Swift} and {\it XMM}, we have investigated the optical/UV variability of Fairall~9, and its connection to the X-ray band, over a range of time scales from hours to months.  Our {\it Swift} monitoring finds significant correlated variability in all optical/UV bands with an amplitude that increases towards shorter wavelengths. We also find a correlation between the UV and X-ray light curves, although it is clear that the X-rays contain a significant component that is not coherent with the UV-band.   In one particularly interesting event, the UW2-band flux is seen to dip by 20\% between two pointings separated by just 4 days, and then recover back to its original level 4-days later.  This dip is not seen in the optical filters or the total X-ray flux, although there is marginal evidence for an X-ray spectral hardening at the time of the dip.  By considering characteristic time scales, we suggest that some significant fraction of the UV emission is reprocessed energy from the central regions of the disk and that the dip corresponds to a temporary shielding of the UV-emission region from the central radiation field.  

On the shorter time scales probed by the {\it XMM} observations, we find rapid (10\,ks) and low-amplitude (2--4\%) UV-flares that may be associated with micro-flares seen in the X-ray flux.  We use arguments based on the blackbody limit and causality to deduce that the region responsible for these flares must have a temperature of at least $8\times 10^4$\,K.  We suggest that these UV-flares correspond to the heating of the centralmost parts of the disk by X-ray flares, although these possibility requires either a large EUV soft excess associated with the X-ray flares or strong beaming of the X-ray flare emission towards the accretion disk.



\acknowledgments
\section*{Acknowledgments}
We thank the referee for his/her useful comments, which led an improvement of the paper. 
This research has made use of: data obtained from the High Energy Astrophysics Science Archive Research Center (HEASARC) provided by NASA's Goddard Space Flight Center. The paper is based on observations obtained with \textit{XMM-Newton}, an ESA science mission with instruments and contributions directly funded by ESA Member States and NASA.


\end{document}